\documentclass[aps,amsfonts,nofootinbib,twocolumn]{revtex4}
\usepackage{epsfig}
\usepackage{graphicx}
\usepackage{amsmath}
\begin{document}
\newcommand{\bb}{\begin{equation}}
\newcommand{\ee}{\end{equation}}
\newcommand{\eqb}{\begin{eqnarray}}
\newcommand{\eqf}{\end{eqnarray}}
\newcommand{\1}{\'{\i}}
\preprint{}
\title{ Electrodynamics with an Infrared Scale and PVLAS
experiment}
\author{Paola Arias}
\email{paola.arias@gmail.com}
\affiliation{Departamento de F\'{\i}sica, Universidad de Santiago de Chile, Casilla 307, Santiago, Chile}
\author{H. Fanchiotti}
\email{huner@fisica.unlp.edu.ar} \affiliation{IFLP/CONICET -
Departamento de F{\'\i}sica, Facultad de Ciencias Exactas,
Universidad Nacional de la Plata, C.C. 67, (1900) La Plata,
Argentina}
\author{J. Gamboa}
\email{jgamboa@usach.cl}
\affiliation{Departamento de F\'{\i}sica, Universidad de Santiago de Chile, Casilla 307, Santiago, Chile }
\author{C. A. Garc{\'\i}a-Canal}
\email{garcia@fisica.unlp.edu.ar} \affiliation{IFLP/CONICET-
Departamento de F{\'\i}sica, Facultad de Ciencias Exactas,
Universidad Nacional de la Plata, C.C. 67, (1900) La Plata,
Argentina}
\author{F. M\'endez}
\email{fmendez@usach.cl}
\affiliation{Departamento de F\'{\i}sica, Universidad de Santiago de Chile, Casilla 307, Santiago, Chile }
\author{F. A. Schaposnik }\thanks{Associated with CICBA}
\email{fidel@fisica.unlp.edu.ar}\affiliation{IFLP/CONICET - Departamento de
F{\'\i}sica, Facultad de Ciencias Exactas, Universidad Nacional de la
Plata, C.C. 67, (1900) La Plata, Argentina\\
 CEFIMAS-SCA, Ave. Santa Fe 1145, C1059ABF, Buenos Aires, Argentina}

\begin{abstract}
We consider an infrared Lorentz violation in connection with recent
results of PVLAS experiment.
  Our analysis is based in a relation that can be established, under certain conditions,
between an axial-like-particle theory and electrodynamics with an
infrared scale. In the PVLAS case, the conditions imply two
dispersion relations such that the infrared scale $|\vec \theta|$,
the inverse  axion-photon coupling constant  $M^{-1}$ and the
external magnetic field ${\vec B}$ can be connected through the
formula $|{\vec \theta}|= {|\vec B|}/M$. Our analysis,  which only requires a non-dynamical
(auxiliar) axial-like field leads to $|{\vec
\theta}| \leq 5.4 \times 10^{-7}$ meV and   $M^{-1} \sim 1.2 \times
10^{-3}$ GeV $^{-1}$.
 \end{abstract}
\date{\today}
\maketitle
\section{Introduction}
Last year the PVLAS collaboration \cite{PVLAS1} reported that when a
linearly polarized laser light crosses a region where there is a
transverse magnetic field, a tiny rotation of the polarization plane
and a birefringence is observed.  This result prompted much activity
in high energy physics because it is an unexpected signal within the
standard quantum electrodynamics \cite{ronca}.

More recently the PVLAS team \cite{PVLAS2} did an update of the
previous results and although the rotation of the polarization plane
was not reconfirmed,  a background ellipticity was measured implying
a birefringence bound $10^{4}$ more bigger than the standard quantum
electrodynamics prediction \cite{bial,adler,adler1} and, therefore,  an explanation out of
the conventional physics seems still to be necessary.

Several groups have proposed different explanations for the
experiment based in axion-like particles (ALP) \cite{ring},
millicharged particles \cite{mohapa}, chameleon fields \cite{uk}  or
refinements of the previous ones.

In this note we would like to study quantum electrodynamics with an infrared scale
and analyze  the birefringence results in its
framework.
The possible existence of an infrared scale in quantum field theory
has been discussed in the literature from different points of view
\cite{IF},  although in this paper we shall
follow \cite{GL,CFJ, Kost,sasha,CCGM}.

The paper is organized as follow: in section II we present
 the electrodynamics with an infrared scale and how
birefringence emerges; we also  establish in this section  the
conditions under which  the infrared modified electrodynamics
becomes equivalent to the ALP approach. In section III we interpret
the PVLAS experiment in terms of the modified electrodynamics and we
establish bounds for the infrared scale and the axion coupling
constant. Finally in section IV, we give our conclusions and
outlook.

\section{Infrared Modified Electrodynamics}

Following \cite{CCGM,GL} the infrared modified electrodynamics is
defined through the  modified  Poisson brackets, {\it i.e.} instead
of considering

\eqb
\left[A_i({\vec x}), A_j({\vec y})\right] &=& 0, \nonumber
\\
\left[A_i({\vec x}), \pi_j({\vec y}) \right] &=& \delta_{ij} \delta
({\vec x},{\vec y}), \label{1}
\\
\left[\pi_i({\vec x}), \pi_j({\vec y})\right] &=& 0, \nonumber \eqf

one writes the following modified commutation relations
\eqb [A_i({\vec x}), A_j({\vec y})]
&=&\kappa_{ij} \delta ({\vec x},{\vec y}), \nonumber
\\
\left[ A_i({\vec x}), \pi_j({\vec y})\right] &=&\delta_{ij} \delta ({\vec x},{\vec y}), \label{2}
\\
\left[ \pi_i({\vec x}), \pi_j({\vec y}) \right] &=&\theta_{ij} \delta ({\vec x},{\vec y}), \nonumber
\eqf
where $\kappa_{ij}$ and  $\theta_{ij}$ are the most general $3\times 3$
constant antisymmetric matrices. From the above relations one
can see that the canonical dimensions for   $\kappa$ and  $\theta$
 are $\mbox{(energy)}^{-1}$ and  $\mbox{(energy)}^{+1}$
respectively. Therefore, as both scales are introduced as tiny
correction to the
 canonical electrodynamics algebra, they can be identified with an
ultraviolet (UV) and an infrared (IR) scale respectively.

In the IR regime --in which we are interested here-- we choose
$\theta_{ij}= \epsilon_{ijk}\theta_k$ and $\kappa_{ij} = 0$ and then
the modified Maxwell equations read
\eqb
\nabla . ~{\vec B} &=& 0, \nonumber
\\
\nabla \times {\vec E} &=& -\frac{\partial {\vec B}}{\partial t}, \label{4}
\\
\nabla .~ {\vec E} &=& -\, \vec \theta . ~{\vec B}, \nonumber
\\
\nabla \times {\vec B} &=& - {\vec E} \times {\vec \theta} + \frac{\partial {\vec E}}{\partial t}. \nonumber
\eqf

The first two equations in  ({\ref{4}) are the standard ones while the other
two ({\it i.e.} the Gauss and Ampere's laws)
are changed. Actually these last two equations  break explicitly Lorentz invariance
 and they lead to the two dispersion relations
\bb
\omega^2_\pm = {\vec k}^2 + \frac{{\vec \theta}^2}{2} \pm \sqrt{({\vec k}.{\vec \theta})^2 + \frac{1}{4} ({\vec \theta}^2)^2}. \label{
7}
\ee
The breaking of Lorentz invariance  and the two dispersion relations
can be understood by noting   that the modified Maxwell equations
are formally equivalent to the standard ones but in a medium with
$-{\vec \theta} \times {\vec A}$ and  ${\vec \theta} A_0$ playing
the role of polarization and magnetization respectively. In such a
situation, it is   natural to expect Lorentz invariance violation
and birefringence.

It is worth noting that equations ({\ref{4}) can be also derived from
the Lagrangian
\bb
{\cal L} = -\frac{1}{4} F^{\mu \nu}F_{\mu \nu}  -
\frac{1}{2} \theta_\mu \epsilon^{\mu \nu \rho \lambda}
A_\nu \partial_\rho A_\lambda. \label{13}
\ee
by taking $(\theta_\mu) = (0,\vec \theta)$. Such a Lagrangian also arises
in the context of the noncommutative field theories \cite{GL} and, moreover,
it is the at the basis of the study of Lorentz and CPT violation
developed by Carroll et al
\cite{CFJ}, Kostelecky et al \cite{Kost} and others \cite{sasha}.

Interestingly enough, it is not difficult to find a connection
between the electrodynamics with an infrared scale discussed above
and the
  ALP model \cite{ring}. Indeed, let us consider the
Lagrangian for axions coupled to electromagnetism,
 \bb {\cal L}_{ALP} = -\frac{1}{4} F^{\mu
\nu}F_{\mu \nu} + \frac{1}{2}
 (\partial \varphi)^2-\frac{1}{2} m^2 \varphi^2 + \frac{1}{4 M} \varphi ~{
\tilde F}^{\mu \nu} F_{\mu \nu}, \label{8}
\ee
 with  ${\tilde F}^{\mu \nu}= \frac{1}{2}
  \epsilon^{\mu \nu \rho \beta}F_{\rho \beta}$, the scalar $\varphi$
is the axion field and $M^{-1}$ the
axion-photon coupling constant. The corresponding
 equations of motion are

 \eqb
( \Box -m^2)\varphi &=& \frac{1}{4M} {\tilde F}^{\mu \nu} F_{\mu \nu}, \nonumber
\\
\nabla .~ {\vec E} &=&-  \frac{1}{M}\nabla \varphi . {\vec B},  \label{10}
\\
\nabla \times {\vec B} &=&  \frac{\partial {\vec E}}{\partial t} +
\frac{1}{M}\left( {\vec E}\times \nabla \varphi - {\vec B} \frac{\partial \varphi}{ \partial t}\right), \nonumber
\eqf
besides the standard ones, {\it i.e.}  $\nabla.~{\vec B}=0$ and $\nabla \times {\vec E} =- \partial {\vec B}/\partial t$.

 One should note that the last two eqs.  in (\ref{4}) coincide
 with the last ones in (\ref{10}) if one establishes a
 correspondence
 \bb
{\vec  \theta} \leftrightarrow \frac{1}{M} \nabla \varphi. \label{11}
 \ee
It should be stressed  that   $\theta_i$ in the l.h.s. is a constant
parameter
 introduced through the
 modification of the canonical commutation relations (\ref{1}) or,
 what is equivalent, through the addition of a Chern-Simons term
 as in (\ref{13}) while $\varphi$
 in the r.h.s is the dynamical axion field.}

Furthermore, one can connect the Lagrangian (\ref{13}) for
electrodynamics with an infrared scale  with the axion Lagrangian
(\ref{8}) by means of the identity

 $$
 {\tilde F}^{\mu \nu} F_{\mu \nu} =
 2 \partial_\mu \left( \epsilon^{\mu \nu \rho \lambda}
  A_\nu \partial_\rho A_\lambda \right).
 $$
 Then, the term $\varphi {\tilde F}^{\mu \nu} F_{\mu \nu}$
 in (\ref{8}) can be integrated by parts  and written, using
 the connection (\ref{11}), in
 terms of the space-like vector $\theta_\mu$

  $$
 (\partial_\mu \varphi)\epsilon^{\mu \nu \rho \lambda} A_\nu \partial_\rho A_\lambda
 \rightarrow M\theta_\mu \epsilon^{\mu \nu
 \rho \lambda} A_\nu \partial_\rho A_\lambda.
 $$
Components of $\theta_\mu$ must be small (tiny actually);
correspondingly the $\nabla \varphi$  components
 must be small and, therefore, the quadratic $(\nabla \varphi)^2$ term in (\ref{8}) can be disregarded while
 the $m^2 \varphi^2$ corresponds in fact to a constant which can be absorbed through a Lagrangian redefinition
 or as normalization constant in the path integral approach. With all this, the Lagrangians (\ref{8})
 and   (\ref{13}) formally coincide.

Once the connection between Lagrangians (\ref{8}) and  (\ref{13}) is
established,  the effects which in the former
 arise due to a dynamical field  can be seen in the
 later as   produced   by the
infrared parameter, $\vec \theta$. This means that although the
 axion interpretation of the PVLAS results seem to be invalidated by recent experiments \cite{france},
our approach suggests that there is no need of an axion participation
in the birefringence phenomenon reported in PVLAS if
 the infrared parameter $|\vec \theta|$ is considered. We discuss this issue in the
 following section.

\section{Interpreting the PVLAS experiment}

In the PVLAS experiment  a linearly polarized photon beam goes
 trough a region where there is an {external} transverse
magnetic field. A non-vanishing ellipticity is observed, which  can
be attributed to an unusual interaction between photons and the
magnetic field. Assuming this, let us interpret the results in the
context of the infrared modified electrodynamics presented in the
precedent section. To this end, we shall exploit the connection that
we established in the precedent section between this model and the ALP theory.


Let us start considering the equations of motion for ALP in the form

 \eqb
 && \Box \varphi - \frac{1}{M} {\dot {\vec A}}\, . \,{\vec B} =0, \label{14}
 \\
 && \Box {\vec A} + \frac{1}{M}{\dot \varphi} {\vec B} =0, \label{15}
 \eqf
 where ${\vec B}$ is the external magnetic field and the Coulomb gauge have been used.

 From these equations one can derive the following dispersion relations

 \bb
 \omega^2_{\pm} = {\vec k}^2 + \frac{{\vec B}^2}{2M^2} \pm \sqrt{\frac{|{\vec k}|^2| {\vec B}|^2}{M^2} +
 \frac{|{\vec B}|^4}{4M^4}}, \label{dis}
 \ee
 which coincide with (\ref{7}) if one makes the identification
\bb
|{\vec \theta}| \leftrightarrow \frac{|{\vec B}|}{M}. \label{ed}
\ee
However, it should be  emphasized that (\ref{ed}) selects only the
magnitude of the vectors ${\theta}$ and ${\vec B}$ but not the
angles between them.

In summary,  the origin of the $\theta$ modification in the Poisson
brackets (\ref{2}) or in the Lagrangian (\ref{13}) should be traced
back to the introduction of an external magnetic field like that in
PVLAS experiment.

Relation (\ref{ed}) expresses the  IR scale  as a connection between the
external magnetic field and the mass scale and, therefore, its
magnitudes cannot be computed directly. However using  the above
dispersion relations, one can find explicit expressions for two different
refractive indices giving rise to birefringence. Indeed, following \cite{maiani}, we chose
${\vec k}\,.\,{\vec \theta}=0$ and then  (\ref{7}) becomes
\bb \omega_+ =
|{\vec k}|, ~~~~~~~\omega_-= \sqrt{|{\vec k}|^2 + |{\vec \theta}|^2},
\label{inf}
\ee
so that the refractive indices are given by
 \bb n_+
= 1, ~~~~~~~ n_- = \frac{ |{\vec k}|}{\sqrt{|{\vec k}|^2 + |{\vec
\theta}|^2}} \approx 1 - \frac{|{\vec \theta}|^2}{2 |{\vec k}|^2}.
\label{refr}
\ee

Consequently, the difference of refractive indices $\Delta n = |n_+ -n_-|$ results in
\begin{equation}
\Delta n = \frac{|{\vec \theta}|^2}{2 |{\vec k}|^2}, \label{i1}
\end{equation}
Since in the infrared modified quantum electrodynamics  $|{\vec \theta}|$ sets the energy scale
at which Lorentz invariance could be violated, these relations show that no violation ($|{\vec \theta}|=0$)
corresponds, in terms of  the external magnetic field, to $\vec B = 0$.

Following the alternative route given by the ALP model one should have
\begin{equation}
\Delta n = \frac{|{\vec B}|^2}{2 M^2|{\vec k}|^2}. \label{i2}
\end{equation}

Relations (\ref{i1})-(\ref{i2}) are of course independent and can be   used for computing ${\vec \theta}$ and $M$
 separately.
Indeed, from the PVLAS data we know that
$$
k\sim 1.2 ~\mbox{eV}, ~~~~~~~~~~~|{\vec B}| \sim 448.5 ~\mbox{eV}^2,
$$
With this and the experimental bound for $\Delta n$, $\Delta n \leq 10^{-19}$, one has
\begin{equation}
|{\vec \theta}| \leq 5.4 \times 10^{-7} ~\mbox{meV},
\label{res1}
\end{equation}
\begin{equation}
M^{-1} \sim  1.2 \times 10^{-3}  ~\mbox{GeV}^{-1}.
\label{res2}
\end{equation}
The value for $M^{-1}$ is three orders above the value obtained by the ALP
model \cite{ring1} and hence the axion does not play, in our approach, any role in
the explanation of the PVLAS experiment. In contrast, the bound for $|{\vec \theta}|$,
  not discussed previously within the PVLAS context, is not excluded by any
  Lorentz violation bound.

\section{Conclusions and outlook}

Quantum electrodynamics calculations, as those  presented in refs.~\cite{adler,adler1,bial},
lead to
 results that are
four orders of magnitude below the birefringence values measured for example in the
PVLAS experiment. This clearly shows that alternative proposals should be
investigated to explain such experiment and in this sense the
possibility of axion-like particle production was an attractive one. However,
as mentioned before, recent ``light shining through a wall'' experiments  \cite{france}
indicate that ALP should be discarded as an explanation for PVLAS
results and hence  new routes different from that of ALP should be investigated
(New experiments in the ALP interpretation will appear the next months,
  see {\it e.g} \cite{ham,otros1,otros2}). In
fact, in the explanation
proposed in this paper, based on a modified version of electrodynamics where an
infrared scale $\vert\vec \theta\vert$ is introduced, possible axial-like particles
are in fact auxiliary and do not take any dynamical role in explaining the PVLAS
experiment.

In our model, birefringence results from the modification of the
dispersion relations produced by the infrared scale $\vert\vec \theta\vert$, which in turn is
connected to the external magnetic field in which the observed
phenomenon takes place. We then conclude that a Lorentz violation such that the bound (\ref{res1})
holds could be at the root of the PVLAS results.

\vspace{0.3 cm}

\noindent\underline{Acknowledgements}:  We would like to thank to O. Bertolami, H. Falomir, E. Fradkin and G. Cantatore by
discussions. This work
was partially supported by FONDECYT-Chile and CONICYT grants 1050114, 1060079 and 21050196, PIP6160-CONICET,
PIC-CNRS/CONICET, BID 1728OC/AR PICT20204-ANPCYT grants and by CIC and UNLP (11/X381 and 11/X450), Argentina.


\begin{thebibliography}{99}
\bibitem{PVLAS1} E. Zavattini et al. [PVLAS collaboration], {\it Phys. Rev. Lett.} {\bf 96}, 110406 (2006).
\bibitem{ronca} See {\it e.g} W. Dittrich and H. Gies, Probing the Quantum Vacuum (Springer, Berlin,
2000); G. V. Dunne,  \lq \lq Heisenberg-Euler effective Lagrangians: Basics and extensions"; M. Roncadelli, Behind PVLAS, [
arXiv:hep-ph/ 0706.4244] and references therein.
\bibitem{PVLAS2} E. Zavattini, et al. [PVLAS collaboration], \lq \lq New PVLAS results and limits on magnetically induced
optical rotation and ellipticity in vacuum", [arXiv:hex/0706.3419].
\bibitem{adler} S. L. Adler, {\it Ann. Phys.} (N.Y.) {\bf 87}, 599 (1971).
\bibitem{bial} Z. Bialynicka-Birula and I. Bialynicka-Birula, {\it Phys. Rev.} {\bf D2}, 2341 (1970).
\bibitem{adler1} S. L. Adler {\it J. Phys.} {\bf A40}, F143 (2007); S. Biswas and K. Melnkov, {\it Phys. Rev.} {\bf D75}, 053003
(2007).
\bibitem{IF} A. G. Cohen, D. B. Kaplan and A. E. Nelson, {\it Phys. Rev. Lett.} {\bf 82}, 4971 (1999); Y.-P. Jing and L. -Z.
Fang, {\it Phys. Rev. Lett.} {\it 73}, 1882 (1994); J. Carmona and J. L. Cortes, {\it Phys. Rev. }{\bf D65}, 025006 (2002); M. V.
Libanov and V. A. Rubakov, {\it JHEP} {\bf 0508}, 001 (2005).
\bibitem{ring} E. Masso and J. Redondo JCAP {\bf 0509}, 015 (2005); J. Jaeckel, E. Masso, J. Redondo, A. Ringwald and F.
Takagashi, \lq \lq We need lab experiments to look for axion-like particles ", hep-ph/0605313;{\it ibid},  \lq \lq The Need for
Purely
Laboratory-Based Axion-Like Particle Searches", hep-ph/0610203 and references therein.
\bibitem{mohapa} R. N. Mohapatra and S. Nasri, hep-ph/0610068, \lq \lq Reconciling the CAST and PVLAS results"; H. Gies,
J. Jaeckel and A. Ringwald, {\it Phys. Rev. Lett.}
{\bf 97}, 140402 (2006).
\bibitem{ring1} {\it e.g} M. Ahlers, H. Gies, J. Jaeckel and A. Ringwald, hep-th/0612098, \lq \lq On the particle interpretation of
the PVLAS data: Neutral versus charged particles"; A. Ringwald, \lq \lq Particle interpretation of the PVLAS data", [hep-ph]
0704.3195.
\bibitem{uk}  P. Brax, C. van Bruck, A. -C. Davis, \lq \lq Testing Chameleon Theories with Light Propagating through a
Magnetic Field \rq \rq, [hep-th] 0707.2801 and references therein.
\bibitem{GL} J. Gamboa and J. L\'opez-Sarri\'on, {\it Phys. Rev.} {\bf D71}, 067702  (2005).
\bibitem{CFJ} S. Carroll, G. Field and R. Jackiw, {\it Phys. Rev.} {\bf D41}, 1231 (1990).
\bibitem{Kost} D.~Colladay and V.~A.~Kostelecky,
  Phys.\ Rev.\  D {\bf 58}, 116002 (1998); R.~Jackiw and V.~A.~Kostelecky,
  Phys.\ Rev.\ Lett.\  {\bf 82}, 3572 (1999); V.~A.~Kostelecky and R.~Lehnert,
  Phys.\ Rev.\  D {\bf 63}, 065008 (2001).
\bibitem{CCGM} J. M. Carmona, J. L. Cort\'es , J. Gamboa and F. M\'endez, JHEP {\bf 0303}, 058 (2003); ibid {\it Phys. Lett.} {
\bf B565}, 222 (2003)
\bibitem{sasha} A. A. Andrianov, R. Soldati and P. Giacconi, JHEP {\bf 0202}, 030 (2002).
\bibitem{maiani} L. Maiani, R. Petronzio and E. Zavattini, {\it Phys. Lett.} {\bf B175}, 359 (1986).
\bibitem{france} C. Robilliard, R. Battesti, M. Fouch\'e, J. Mauchain,A.M. Sautivet and Carlo Rizzo, \lq \lq No light
shinning
through a wall", [hep-ex] 0707.1296.
\bibitem{ham} S. J. Chen, H. H. Mei and W. T. Ni, (Q\& A Collaboration), hep-ex/0611050.
\bibitem{otros1}   C. Rizzo (BMV collaboration), 2nd ILIAS-CERN-CAST axion Academic Training 2006,
http://cast.mppmu.mpg.de.
\bibitem{otros2} P. Pugnat {\it et al}, (OSQAR Collaboration), CERN-SPSC-2006-035, CERN-SPSC-P-331.
\end{thebibliography}
\end{document}